\documentclass[journal,onecolumn, compsoc]{IEEEtran}
\usepackage{amsmath,amsfonts,amssymb}
\usepackage{algorithmic}
\usepackage{algorithm}
\usepackage{array}
\usepackage[caption=false,font=footnotesize,labelfont=rm,textfont=sf]{subfig}
\usepackage{textcomp}
\usepackage{xcolor}
\usepackage{url}
\usepackage{verbatim}
\usepackage{graphicx}
\usepackage{cite}
\usepackage{bm}
\usepackage{tikz}
\graphicspath{{./figure/}}

\newtheorem{Rem}{Remark}

\newtheorem{Thm}{Theorem}
\newtheorem{Cor}{Corollary}

\newcommand{\norm}[1]{\left\lVert#1\right\rVert}

\begin{document}

\title{On the Channel Correlation in Reconfigurable Intelligent Surface-Aided System}

\author{ Kuang-Hao (Stanley) Liu,~\IEEEmembership{Member,~IEEE}
\thanks{K-H.~Liu is with the Institute of Communications Engineering,
National Tsing Hua University, Hsinchu, Taiwan 300044. E-mail: {\tt khliu@ee.nthu.edu.tw}}
}




\maketitle

\begin{abstract}
This works explores the correlation between channels in reconfigurable intelligent surface (RIS)-aided communication systems. In this type of system, an RIS made up of many passive elements with adjustable phases reflects the transmitter's signal to the receiver. Since the transmitter-RIS link may be shared by multiple receivers, the cascade channels of two receivers may experience correlated fading, which can negatively impact system performance. Using the mean correlation coefficient as a metric, we analyze the correlation between two cascade channels and derive an accurate approximation in closed form. We also consider the extreme case of an infinitely large number of RIS elements and obtain a convergence result. Our analysis accuracy is validated by simulation results, which offer insights into the correlation characteristics of RIS-aided fading channels.
\end{abstract}


\section{Introduction}
Recently, reconfigurable intelligent surface (RIS) has been received significant attention as a new wireless communication paradigm~\cite{Wu2019,Tang2020,Liaskos2018,Basar2019,Wu2020}. An RIS is a meta-surface consisting of a large number of low-cost passive reflecting elements. The phase of each element can be electronically tuned to reflect the radio signals in a desired manner. 
In the multi-user scenario, an RIS may be used to assist the base station (BS) for serving multiple users. Since the channel between the BS and RIS is common to all users, the signals reflected from the RIS may experience correlated fading that deteriorates the degree of freedom in the fading channel. It is thus crucial to characterize the channel correlation in RIS-aided communications.


The issue of correlated fading in RIS-aided communications has been investigated in some recent work. Considering the spatial correlation due to collocated RIS elements, \cite{Nadeem2020} studied the channel estimation problem in RIS-aided communications. In~\cite{Chien}, the phase shift design for RIS-aided communications is investigated and it shows that the presence of spatial correlation may simplify the phase configuration. In~\cite{Psomas2021}, the authors theoretically analyzed the impact of spatial correlation between RIS elements to the performance of simultaneous wireless information and power transfer (SWIPT). Theirs results reveal different impacts of spatial correlation to the average harvested energy and outage probability. The work in ~\cite{Papazafeiropoulos2021} also evaluates the performance of spatially correlated RIS-aided communications with multiple RISs. Existing work focuses on the spatial correlation between reflection elements of the RIS. Another level of correlation arises between different RIS-aided channels when the same RIS is used to reflect the signals of different users. In the multi-user setting, the channel correlation is a degrading factor due to the low-rank channel matrix that deteriorates the multiplexing gain.

In this work, we study the correlation between two RIS-aided cascade channels, each consisting of a user-independent channel between the BS and the RIS and a user-specific channel between the RIS and the user. Since each cascade channel suffers different fading, their correlation is evaluated by computing the correlation coefficient averaged over individual fading component. For the BS-RIS channel shared by different users, whether the line-of-sight (LoS) exists may affect the average correlation coefficient differently and thus the user-independent channel is modeled by the Rician channel. On the other hand, the user-specific channel between the RIS and the user is modeled by the Rayleigh channel. The considered channel model captures the practical deployment scenario where the passive RIS is installed in a spot with a clear view to the BS while users are likely to be surrounded by dense scatters without a LoS. Based on the considered channel model, we derive the mean correlation coefficient of two cascade RIS channels in closed-form. The obtained expression allows us to identify the key factor that dominates the channel correlation and the asymptotic behavior when the number of RIS elements is large. A related issue is the channel hardening effect of RIS, which concerns how many RIS elements are required to mitigate channel variations due to fading for a single user~\cite{Bjoernson2021}. Differently, our analysis provides a direct evidence on how fast the correlation between two users associated with a common RIS decays with the number of BS antennas and RIS elements, respectively. Our work can be extended to the Nakagami fading distribution yet the obtained convergence result does not change with the underlying fading model. This can be justified because the Nakagami model behaves approximately the same as the Rician model near their mean values. Also, key results drawn from this work hold for any given phase-shift configuration irrelevant to the underlying optimization algorithms for the phase-shift design. 

The remainder of the paper is organized as follows. Sec.~\ref{sec: system model} explains the considered RIS system and channel models. To facilitate theoretical analysis, some preliminary results are given in Sec.~\ref{sec: preliminary} followed by the main results in Sec.~\ref{sec: analysis}. Numerical results are discussed in Sec.~\ref{sec: results}. Finally, Sec.~\ref{sec: conclusion} concludes this work.

\emph{Notations}: In our notations, italic letters are used for scalars. Vectors and matrices are noted by bold-face letters. For a complex-valued vector $\mathbf{x}$, $\norm{\mathbf{x}}$ denotes its Euclidean norm. If $\mathbf{x}$ and $\mathbf{y}$ are two complex-valued vectors of the same dimension, their inner product is represented as $\mathbf{x}\cdot \mathbf{y}^H$. For a complex-valued scalar $a$, $a^*$ denotes the conjugate of $a$. For a matrix $\mathbf{A}$, $\mathbf{A}^T$ and $\mathbf{A}^H$ denote the transpose and conjugate transpose, respectively. $\mathbf{I}_N$ denotes an $N\times N$ identity matrix.  $\mathbb{E}[\cdot]$ denotes the statistical expectation. Finally, $\mathbb{C}^{m \times n}$ denotes the space of $m\times n$ complex-valued matrices.

\section{System Model}\label{sec: system model}

\begin{figure}[!t]
\centering
{
\includegraphics[width=0.75\linewidth]{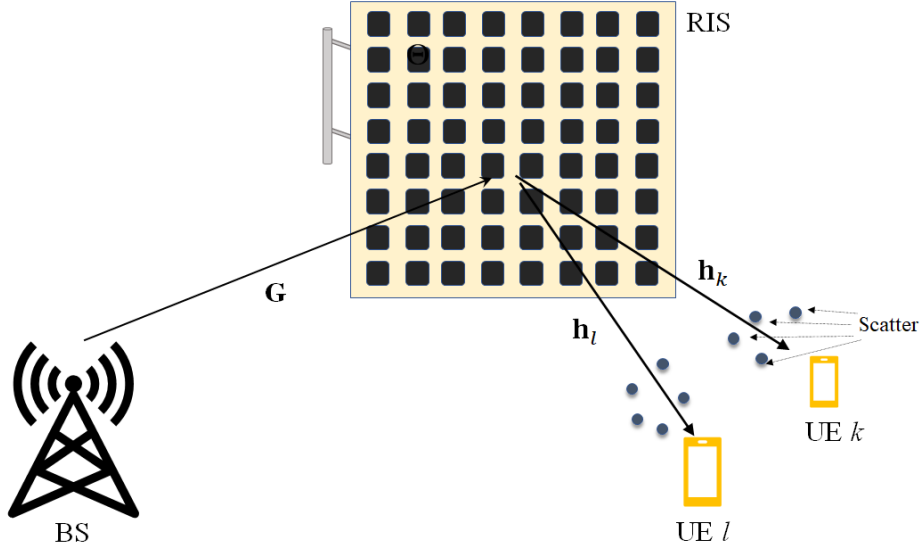}
}
\caption{System configurable of a RIS-aided system.}
\label{fig: system-model}
\end{figure}

As shown in Fig.~\ref{fig: system-model}, we consider an RIS-assisted communication scenario, where an RIS is deployed between the base station (BS) and user equipments (UEs). The BS has $M$ active antennas in uniform linear array (ULA) and the RIS has $N$ reflective elements in uniform planar array (UPA)~\cite{Shen2021}. Each UE has a single antenna. The direct link between the BS and the UE is ignored due to severe propagation loss or blockage. At the downlink, the received signal at the $k$-UE can be expressed as
\begin{equation}\label{eq: signal-model}
y_k = \mathbf{h}_{k} \boldsymbol{\Theta} \mathbf{G} \mathbf{x}+n_k
\end{equation}
where $\mathbf{x}$ is the transmitted signal, $\mathbf{h}_k \in \mathbb{C}^{1\times N}$ and $\mathbf{G} \in \mathbb{C}^{N \times M}$ denote the RIS-UE and the BS-RIS channels, respectively; $n_k$ represents the noise term; $\boldsymbol{\Theta}=\text{diag}(e^{j\varphi_1}, \cdots, e^{j\varphi_N})$ is a diagnomal matrix where $\varphi_n \in [0, 2\pi]$ is the phase shift of the $n$th reflective element on the RIS. From \eqref{eq: signal-model}, the cascade channel from the BS to the $k$-th UE through the RIS can be written as
\begin{equation}\label{eq: cascade-channel}
\mathbf{h}_{k} \boldsymbol{\Theta} \mathbf{G} \triangleq \mathbf{H}_k.
\end{equation}
Since the RIS is passive in nature, deploying the RIS with line-of-sight (LoS) is beneficial to the reflected signal strength~\cite{Wu2019}. Thus, the BS-RIS link is modeled as the Rician fading channel given by
\begin{equation}
\mathbf{G} =  \sqrt{\frac{\kappa}{\kappa+1}} \mathbf{G}_{\text{LOS}} + \sqrt{\frac{1}{\kappa+1}} \mathbf{G}_{\text{NLOS}}
\end{equation}
where $\kappa$ denotes the Rician factor, $\mathbf{G}_{\text{LOS}}$ and $\mathbf{G}_{\text{NLOS}}$ the LoS and the Non-LoS (NLoS) components, respectively. Let $\psi_1\in (-\pi/2,\pi/2)$ and $\theta_1\in (-\pi/2,\pi/2)$ denote the azimuth and elevation angle of arrival (AoA) of the LoS path of the BS-RIS channel, respectively. For the UPA with $N_1$ horizontal elements and $N_2$ vertical elements, the $N_1\cdot N_2 \times 1$ steering vector of the BS-RIS channel is given by
\begin{equation}\label{eq: b_R}
\mathbf{b}_R(\psi_1,\theta_1) = \frac{1}{\sqrt{N_1}}\left[ e^{j2\pi n_1 \frac{d_R}{\lambda}\cos(\theta_1)\sin(\psi_1)} \right]^T_{n_1\in \mathcal{I}(N_1)} \otimes \frac{1}{\sqrt{N_2}}\left[ e^{j2\pi n_2 \frac{d_R}{\lambda}\sin(\theta_1)} \right]^T_{n_2\in \mathcal{I}(N_2)}
\end{equation}
where $d_R$ is the distance between two neighboring RIS elements, $\lambda$ is the wavelength, $\mathcal{I}(n)=\{0,\cdots,n-1\}$. As to the ULA at the BS, the steering vector of the BS-RIS channel is given by
\begin{equation}\label{eq: a_T}
\mathbf{a}_T(\psi^{\text{AoD}}) = \frac{1}{\sqrt{M}}\left[ e^{j2\pi n_1 \frac{d_B}{\lambda}\sin(\psi^{\text{AoD}})} \right]^T_{m\in \mathcal{I}(M)} 
\end{equation}
where $\psi^{\text{AoD}}\in (-\pi/2,\pi/2)$ denotes the angle of departure (AoD) and $d_B$ represents the antenna spacing. Then the LoS path between the BS and the RIS can be modeled as
\begin{equation}\label{eq: G_los}
\mathbf{G}_{\text{LOS}} = \mathbf{b}_R(\psi_1,\theta_1) \mathbf{a}^H_T(\psi^{\text{AoD}}).
\end{equation}
On the other hand, $\mathbf{G}_{\text{NLOS}}$ is an $N\times M$ matrix where each entry follows the complex normal distribution with zero mean and unit variance. To model the user-specific RIS-UE channel, the spatial-domain model is considered as given by
\begin{equation}\label{eq: ris-user channel}
\mathbf{h}_{k} = \sum_{p=1}^{L_{k}} \alpha_{p,k} \mathbf{b}_T^H(\psi_{2,k,p},\theta_{2,k,p}) 
\end{equation}
where $L_{k}$ and $\alpha_{p,k}$ denote the number of resolvable paths of the $k$-th RIS-UE channel and the complex gain of the $p$-th path with independent and identically distributed (i.i.d) complex Gaussian entries; $\psi_{2,k,p}\in (-\pi/2,\pi/2)$ and $\theta_{2,k,p}\in (-\pi/2,\pi/2)$ are the azimuth and elevation AoDs of the $p$-th path, respectively, and $\mathbf{b}_T(\psi_{2,k,p},\theta_{2,k,p})$ is the $N\times 1$ steering vector given by
\begin{equation}\label{eq: b_T}
\mathbf{b}_T(\psi_{2,k,p},\theta_{2,k,p}) = \frac{1}{\sqrt{N}} \left[ e^{j2\pi n_1 \frac{d_R}{\lambda}\cos(\theta_{2,k,p})\sin(\psi_{2,k,p})} \right]^T_{n_1\in \mathcal{I}(N_1)} \otimes \frac{1}{\sqrt{N}} \left[ e^{j2\pi n_2 \frac{d_R}{\lambda}\sin(\theta_{2,k,p})} \right]^T_{n_2\in \mathcal{I}(N_2)}.
\end{equation}

%

\section{Preliminary}\label{sec: preliminary}

To facilitate our analysis, it is convenient to express the spatial channel models in terms of the normalized spatial frequencies. Let $\phi_i \triangleq \frac{d_R}{\lambda} \cos(\theta_i)\sin(\psi_i) \in (-d_R/\lambda,d_R/\lambda)$ and $\vartheta_i \triangleq \frac{d_R}{\lambda} \sin(\theta_i)\in (-d_R/\lambda,d_R/\lambda)$ for $i=1,2$. Then the steering vectors in \eqref{eq: b_R} and \eqref{eq: b_T} can be alternately represented as
\begin{align}\label{eq: b_R_-b_T-V2}
\mathbf{b}_R(\psi_1,\theta_1) &= \frac{1}{\sqrt{N}}\mathbf{b}(\phi_1,\vartheta_1), \nonumber \\
\mathbf{b}_T(\psi_{2,k,p},\theta_{2,k,p}) &= \frac{1}{\sqrt{N}}\mathbf{b}(\phi_{2,k,p},\vartheta_{2,k,p})
\end{align}
where $\mathbf{b}(\phi,\theta)$ is defined as
\begin{align}\label{eq: steering-b}
\mathbf{b}(\phi,\vartheta) = [e^{jb_{n_1\cdot N_2+n_2}(\phi,\vartheta)}]^T_{n_1\in \mathcal{I}(N_1), n_2\in \mathcal{I}(N_2)}
\end{align}
with $b_{n_1\cdot N_2+n_2}(\phi,\vartheta) = 2\pi(n_1\phi+n_2 \vartheta)$. With the normalized spatial frequency $\phi^{\text{AoD}} \triangleq \frac{d_B}{\lambda} \sin(\psi^\text{AoD})\in (-d_B/\lambda,d_B/\lambda)$, \eqref{eq: a_T} can be represented as
\begin{equation}\label{eq: steering-a}
\mathbf{a}(\phi^{\text{AoD}}) = \frac{1}{\sqrt{M}}[e^{j 2\pi m \phi^{\text{AoD}}}]^T_{m\in \mathcal{I}(M)} .
\end{equation}

Since the BS-RIS channel is the superposition of the LoS and NLoS components with different fading statistics, $\mathbf{H}_k$ can be represented as
\begin{equation}\label{eq: channel-structure}
\mathbf{H}_k = \mathbf{H}_{k,\text{LOS}} + \mathbf{H}_{k,\text{NLOS}}
\end{equation}
where 
\begin{subequations}
\begin{align}
\mathbf{H}_{k,\text{LOS}} &= \sqrt{\frac{\kappa}{\kappa+1}}\bm{p}^T \text{diag}(\mathbf{h}_k^T) \mathbf{G}_\text{LOS} \label{eq: H_los} \\
\mathbf{H}_{k,\text{NLOS}} &= \sqrt{\frac{1}{\kappa+1}}\bm{p}^T \text{diag}(\mathbf{h}_k^T) \mathbf{G}_\text{NLOS} \label{eq: H_nlos}
\end{align}
\end{subequations}
with $\bm{p}$ being the reflecting coefficient vector of the RIS given by $\bm{p}^T=[e^{j\varphi_n}]^T_{n \in \mathcal{I}(N)}$. \eqref{eq: H_los} and \eqref{eq: H_nlos} are obtained from rewriting \eqref{eq: cascade-channel} based on the property of the diagonal matrix. Using \eqref{eq: b_R_-b_T-V2} and \eqref{eq: steering-a}, \eqref{eq: H_los} can be rewritten as
\begin{equation}\label{eq: Hk_los}
\mathbf{H}_{k,\text{LOS}} = \frac{1}{N}\sqrt{ \frac{\kappa}{\kappa+1} } \sum_{p=1}^{L_k}  \alpha_{k,p} \bm{p}^T \mathbf{b}(\phi_1-\phi_{2,k,p}, \vartheta_1-\vartheta_{2,k,p}) \mathbf{a}^H(\phi^\text{AoD}) 
\end{equation}
where $\bm{p}^T \mathbf{b}(\phi_1-\phi_{2,k,p},\vartheta_1-\vartheta_{2,k,p})$ is a complex coefficient given by
\begin{align}\label{eq: p^Tb}
\bm{p}^T  \mathbf{b}(\phi_1-\phi_{2,k,p},\vartheta_1-\vartheta_{2,k,p}) = \sum_{n_1=0}^{N_1-1} \sum_{n_2=0}^{N_2-1} \chi_{k,p,n_1\cdot N_2+n_2}
\end{align}
where $\chi_{k,p,n_1\cdot N_2+n_2} = e^{j [\varphi_{n_1\cdot N_2+n_2+1}+b_{n_1\cdot N_2+n_2}(\phi,\vartheta)]}$. Then $\mathbf{H}_{k,\text{LOS}}$ can be expressed as
\begin{equation}\label{eq: Hk_los_v2}
\mathbf{H}_{k,\text{LOS}} = \frac{1}{N}\sqrt{ \frac{\kappa}{\kappa+1} } \sum_{p=1}^{L_k}  \alpha_{k,p} \sum_{n_1=0}^{N_1-1} \sum_{n_2=0}^{N_2-1} \chi_{k,p,n_1\cdot N_2+n_2} \mathbf{a}^H(\phi^\text{AoD}). 
\end{equation} 
Similarly, the user-specific channel $\mathbf{h}_k$ can be represented as
\begin{equation}\label{eq: ris-user channel-v2}
\mathbf{h}_{k} = \frac{1}{\sqrt{N}}\sum_{p=1}^{L_{k}} \alpha_{p,k} \mathbf{b}^H(\phi_{2,k,p},\vartheta_{2,k,p}) 
\end{equation}

\begin{Rem}
The exact value of $\bm{p}^T  \mathbf{b}(\phi_1-\phi_{2,k,p},\vartheta_1-\vartheta_{2,k,p})$ depends on how the RIS elements are tuned for the given BS-RIS channel $\mathbf{G}$ and the RIS-UE channel $\mathbf{h}_k$. To boost the received signal quality, each entry in the reflecting coefficient vector $\bm{p}$ should be adjusted constructively to compensate the phase offset of the received signal due to channel fading of the BS-RIS and the RIS-UE channels. Alternatively, it can be tuned destructively to suppress interference~\cite{Wu2019} for the intended signal or hide private information from the eavesdropper~\cite{Hsu2022}. Without loss of generality, the following analysis is applied to any given RIS reflecting coefficient configuration. 
\end{Rem}

The mean power of the small-scale fading coefficients will be frequently encountered in the analysis. For the $k$-th UE, different paths of the RIS-UE channel experience independent fading, and thus
\begin{equation}\label{eq: alpha_square}
\mathbb{E}[\alpha_{k,p} \alpha^*_{k,p'}] =\begin{cases}
1, & p=p' \\
0, & p\neq p'.
\end{cases}
\end{equation}
Similarly, different RIS elements suffer independent fading in the BS-RIS channel. Hence,
\begin{equation}\label{eq: g_square}
\mathbb{E}[g_{n,m} g^*_{n',m}] =\begin{cases}
1, & n=n' \\
0, & n\neq n',
\end{cases}
\end{equation}
where $g_{n,m}$ represents the $(n,m)$-th entry of $\mathbf{G}_{\text{NLOS}}$. On the other hand, $\phi_1$, $\phi_{2,k,p}$, $\vartheta_{2,k,p}$ are independent and uniformly distributed in $(-d_R/\lambda,d_R/\lambda)$. Consequently, 
\begin{equation}\label{eq: chi}
\mathbb{E}[\chi_{k,p,n} \chi^*_{k,p,n'}]=\begin{cases}
1, & n=n' \\
0, & n \neq n'.
\end{cases}
\end{equation}
With \eqref{eq: chi}, it is readily obtained that
\begin{equation}\label{eq: |p^Tb|^2}
\mathbb{E}[ |\bm{p}^T  \mathbf{b}(\phi_1-\phi_{2,k,p},\vartheta_1-\vartheta_{2,k,p})|^2] = N.
\end{equation}

\section{Analysis of Mean Correlation Coefficient}\label{sec: analysis}
The correlation coefficient between two channel vectors $\mathbf{H}_k$ and $\mathbf{H}_l$ has been defined as~\cite{Hoydis2012}
\begin{align}\label{eq: corr-coeff-def}
\rho_{k,l} = \frac{|\mathbf{H}_{k} \cdot \mathbf{H}_{l}^H|}{\Vert \mathbf{H}_{k} \Vert \norm{\mathbf{H}_{l}}}.
\end{align} 
%
To find the mean of $\rho_{k,l}$, the joint distribution of the user-independent channel $\mathbf{G}$ and the user-specific channels $\mathbf{h}_{k}$ and $\mathbf{h}_{l}$ is required. However, it is difficult to obtain the exact joint distribution because $\mathbf{G}$, $\mathbf{h}_{k}$ and $\mathbf{h}_{l}$ are random matrix (vectors) with non-zero means. Besides, the numerator and the denominator in \eqref{eq: corr-coeff-def} contains the common channel $\mathbf{G}$ and thus they are correlated. In the analysis, we ignore the dependence between the numerator and the denominator to obtain an approximate for $\rho_{k,l}$ by taking the expectation toward the numerator and the denominator separately. The accuracy of this approximation will be examined through numerical results. Furthermore, it is easier to find the mean of $|\mathbf{H}_k \cdot \mathbf{H}_l^H|^2$ than $|\mathbf{H}_k \cdot \mathbf{H}_l^H|$ in the numerator of \eqref{eq: corr-coeff-def} since $|\mathbf{H}_k \cdot \mathbf{H}_l^H|^2$ can be computed  as $(\mathbf{H}_k \cdot \mathbf{H}_l^H)( \mathbf{H}_l \cdot \mathbf{H}_k^H)$. Accordingly, the denominator in \eqref{eq: corr-coeff-def} should replaced by $\| \mathbf{H}_k \|^2 \| \mathbf{H}_l \|^2$ but its expectation is difficult to be evaluated because $\mathbf{H}_k$ and $\mathbf{H}_l$ share the same BS-RIS channel and thus they are correlated. Again, the dependence is ignored, leading to the following approximation for the mean squared correlation coefficient.
\begin{subequations}
\begin{align}
\mathbb{E}[\rho^2_{i,j}] &= \mathbb{E} \left[ \frac{ |\mathbf{H}_{k} \cdot \mathbf{H}_{l}^H|^2}{ \| \mathbf{H}_{k} \|^2 \| \mathbf{H}_{l} \|^2} \right]   \label{eq: approx-mean-squared-corr-coeff-dependence} \\
&\approx  \frac{ \mathbb{E} \left[ |\mathbf{H}_{k} \cdot \mathbf{H}_l^H|^2  \right]}{ \mathbb{E}\left[ \Vert \mathbf{H}_{k} \Vert^2 \right] \mathbb{E} \left[ \Vert \mathbf{H}_{l}  \Vert^2 \right] }. \label{eq: approx-mean-squared-corr-coeff-indep}
\end{align}
\end{subequations}
Since the square root function is concave, the desired mean correlation coefficient can be obtained using Jensen's inequality as
\begin{equation}\label{eq: ub-mean-corr-coeff}
\mathbb{E}[ \rho_{i,j} ] = \mathbb{E}\left[ \sqrt{\rho^2_{i,j}} \right] \leq \sqrt{ \mathbb{E}[ \rho^2_{i,j} ] }
\end{equation}
In the sequel, we derive~\eqref{eq: approx-mean-squared-corr-coeff-indep} and start from the mean channel power $\mathbb{E}[\|\mathbf{H}_k\|^2]$.

\subsection{Mean Power of One Cascade Channel}
From \eqref{eq: channel-structure}, each cascade channel contains the contributions from the LOS and NLOS components. Hence, the mean power of the cascade channel per user can be given by
\begin{equation}\label{eq: H-square}
\mathbb{E}[\|\mathbf{H}_k\|^2] = \mathbb{E}[ \|\mathbf{H}_{k,\text{LOS}} \|^2 ] + \mathbb{E}[ \|\mathbf{H}_{k,\text{NLOS}} \|^2 ] + \mathbb{E}[\mathbf{H}_{k,\text{LOS}} \cdot \mathbf{H}^H_{k,\text{NLOS}}] + \mathbb{E}[\mathbf{H}_{k,\text{NLOS}} \cdot \mathbf{H}^H_{k,\text{LOS}}].
\end{equation}
Each term in the right-hand side of \eqref{eq: H-square} can be obtained as follows.

\begin{Thm}\label{Thm: 1}
The mean power of the LOS components in the cascade RIS channel is given by 
\begin{align}\label{eq: H_los_square_final}
\mathbb{E}[\| \mathbf{H}_{k, \text{LOS}} \|^2] = \frac{L_k}{N} \frac{\kappa}{\kappa+1}.
\end{align}
\end{Thm}
\begin{IEEEproof}
From~\eqref{eq: Hk_los} and because $\mathbf{a}^H(\phi^{\text{AoD}}) \cdot \mathbf{a}(\phi^{\text{AoD}})=1$, the mean power of the LOS component can be obtained as
\begin{align}\label{eq: H_los_square_v2}
\mathbb{E}[\| \mathbf{H}_{k,\text{LOS}} \|^2] 
&= \frac{\kappa}{\kappa+1} \frac{1}{N^2} \sum_{p=1}^{L_k} \mathbb{E}[ |\bm{p}^T \mathbf{b}|^2 ].
\end{align}
Then using \eqref{eq: |p^Tb|^2}, we obtain \eqref{eq: H_los_square_final}.
\end{IEEEproof}

\begin{Thm}\label{Thm: 2}
The mean power of the NLOS components in the cascade RIS channel is given by 
\begin{align}\label{eq: H_nlos_square_final}
\mathbb{E}[\| \mathbf{H}_{k, \text{NLOS}} \|^2] = \frac{M L_k}{\kappa+1}.
\end{align}
\end{Thm}
\begin{IEEEproof}
According to \eqref{eq: H_nlos}, the mean power of the NLOS equivalent channel is given by 
\begin{equation}\label{eq: eq: H_nlos_square_v1}
\mathbb{E}[\|\mathbf{H}_{k,\text{NLOS}}\|^2] = \frac{1}{\kappa+1}\bm{p}^T \mathbb{E}_{ \mathbf{h}_k, \mathbf{G}_{\text{NLOS}}} \left[\text{diag}(\mathbf{h}_k^T)\mathbf{G}_{\text{NLOS}}\mathbf{G}^H_{\text{NLOS}} \text{diag}(\mathbf{h}_k^*)\right] \bm{p}^*.
\end{equation}
The independence between $\mathbf{h}_k$ and $\mathbf{G}_{\text{NLOS}}$ suggests that we can find the expectation of $\mathbf{G}_{\text{NLOS}} \mathbf{G}^H_{\text{NLOS}}$ first. Since each entry in $\mathbf{G}_{\text{NLOS}}$ is complex Gaussian distributed with zero and unit variance, we have
\begin{equation}\label{eq: mean_G_nlos}
\mathbb{E}[ \mathbf{G}_{\text{NLOS}}\mathbf{G}^H_{\text{NLOS}}] = M\mathbf{I}_N.
\end{equation}
Next, $\text{diag}(\mathbf{h}_k^T) \text{diag}(\mathbf{h}_k^*)$ remains a diagonal matrix and thus its expectation can be obtained as
\begin{align}
\mathbf{E}[ \text{diag}(\mathbf{h}_k^T) \text{diag}(\mathbf{h}_k^*)] &= \frac{1}{N} \text{diag}\Bigl( \sum_{p=1}^{L_k} \sum_{p'=1}^{L_k} \mathbb{E}[\alpha_{k,p} \alpha_{k,p'}^*] \cdot \mathbb{E}[e^{j b_0(\phi_{k,2,p}-\phi_{k,2,p'}, \vartheta_{k,2,p}-\vartheta_{k,2,p'})}], \nonumber \\
& \cdots, \sum_{p=1}^{L_k} \sum_{p'=1}^{L_k} \mathbb{E}[\alpha_{k,p} \alpha_{k,p'}^*] \cdot \mathbb{E}[ e^{j b_{N-1}(\phi_{k,2,p}-\phi_{k,2,p'}, \vartheta_{k,2,p}-\vartheta_{k,2,p'})}]\Bigr)
\end{align}
where $b_n$ for $n=0,\cdots,N-1$ is defined in \eqref{eq: steering-b}. According to \eqref{eq: alpha_square}, $\mathbb{E}[\alpha_{k,p} \alpha_{k,p'}^*]$ is equal to one when $p=p'$, in which case $e^{jb_n(\phi_{k,2,p}-\phi_{k,2,p'}, \vartheta_{k,2,p}-\vartheta_{k,2,p'})}=1$. For $p \neq p'$, $\mathbb{E}[\alpha_{k,p} \alpha_{k,p'}^*]$ equals zero. Thus, 
\begin{equation}\label{eq: mean_diag_h}
\mathbf{E}[ \text{diag}(\mathbf{h}_k^T) \text{diag}(\mathbf{h}_k^*)] = \frac{L_k}{N} \mathbf{I}_N 
\end{equation}
Substituting \eqref{eq: mean_G_nlos} and \eqref{eq: mean_diag_h} into \eqref{eq: eq: H_nlos_square_v1}, we obtain
\begin{align}
\mathbb{E}[\|\mathbf{H}_{k,\text{NLOS}}\|^2] &= \frac{ML_k}{N(\kappa+1)} \bm{p}^T \mathbf{I}_N \bm{p}^* \nonumber \\
&=  \frac{ML_k}{\kappa+1}
\end{align}
using the fact that $\bm{p}^T \bm{p}^*=N$.
\end{IEEEproof}


\begin{Thm}\label{Thm: 3}
The expectations of the inner product, $\mathbb{E}[\mathbf{H}_{k,\text{LOS}} \cdot \mathbf{H}_{k,\text{NLOS}}^H]$ and $\mathbb{E}[\mathbf{H}_{k,\text{NLOS}} \cdot \mathbf{H}_{k,\text{LOS}}^H]$, both are equal to zero.
\end{Thm}
\begin{IEEEproof} 
Recall that both $\mathbf{H}_{k,\text{LOS}}$ and $\mathbf{H}_{k,\text{NLOS}}$ are $1\times M$ vectors. Based on \eqref{eq: Hk_los_v2}, the $m$-th entry of $\mathbf{H}_{k,\text{LOS}}$ can be expressed as
\begin{equation}
\mathbf{H}_{k,\text{LOS}}(m) =\frac{1}{N} \sqrt{\frac{\kappa}{\kappa+1}} \sum_{p=1}^{L_k} \sum_{n_1=0}^{N_1-1} \sum_{n_2=0}^{N_2-1}  \alpha_{k,p} \chi_{k,p,n_1\cdot N_2+n_2} e^{-j2\pi (m-1) \phi^{\text{AoD}} }
\end{equation}
for $m=1,\cdots,M$. Based on \eqref{eq: H_nlos}, the $m$th entry of $\mathbf{H}_{k,\text{NLOS}}$ can be expressed as
\begin{equation}
\mathbf{H}_{k,\text{NLOS}}(m) = \sqrt{\frac{1}{\kappa+1}} \sum_{n=1}^N g_{n,m} \sum_{p=1}^{L_k} \alpha_{k,p} e^{j (\varphi_n+b_{n+1})}.
\end{equation}
Then the product $\mathbf{H}_{k,\text{LOS}}(m) \mathbf{H}^*_{k,\text{NLOS}}(m)$ is given by
\begin{equation}\label{eq: inner-product_Hlos_Hnlos}
\mathbf{H}_{k,\text{LOS}}(m) \mathbf{H}^*_{k,\text{NLOS}}(m) = \frac{1}{N} \sqrt{\frac{\kappa}{(\kappa+1)^2}} \sum_{p=1}^{L_k} \sum_{n_1=0}^{N_1-1} \sum_{n_2=0}^{N_2-1} \sum_{n'=1}^N \sum_{p'=1}^{L_k}  \alpha_{k,p} \alpha_{k,p'} g^*_{n',m} \chi_{k,p,n_1\cdot N_2+n_2} e^{-j[2\pi (m-1) \phi^{\text{AoD}}+\varphi_{n'}+b_{n'+1} ]}.
\end{equation}
Since $g_{n',m}$ is independent from other variables in~\eqref{eq: inner-product_Hlos_Hnlos} and its mean is zero, $\mathbb{E}[\mathbf{H}_{k,\text{LOS}}(m)\cdot \mathbf{H}^*_{k,\text{NLOS}}(m)]$ is equal to zero. Together with the fact that $\mathbf{H}_{k,\text{LOS}} \cdot \mathbf{H}^H_{k,\text{NLOS}} = \sum_{m=1}^M \mathbf{H}_{k,\text{LOS}}(m) \mathbf{H}^*_{k,\text{NLOS}}(m)$, we obtain $\mathbb{E}[\mathbf{H} _{k,\text{LOS}}\mathbf{H}^H_{k,\text{NLOS}}]=0$. Likewise, it can be shown that $\mathbb{E}[\mathbf{H}_{k,\text{NLOS}} \cdot \mathbf{H}^H_{k,\text{LOS}}]=0$.
\end{IEEEproof}

\subsection{Mean Power of the Inner Product between Two Cascade Channels}

Next, we derive the mean power of the inner product between two cascade channels, i.e., $\mathbb{E} \left[ |\mathbf{H}_{k} \cdot \mathbf{H}_l^H|^2  \right]$ appeared in \eqref{eq: approx-mean-squared-corr-coeff-indep}. Firstly, $|\mathbf{H}_{k} \cdot \mathbf{H}_l^H|^2$ can be expressed as
\begin{equation}\label{eq: HkHl-power}
|\mathbf{H}_{k} \cdot \mathbf{H}_l^H|^2 = \mathbf{H}_k \mathbf{H}_l^H \mathbf{H}_l \mathbf{H}_k^H
\end{equation}
 Since each cascade channel contains the LOS and NLOS parts with different fading statistics, $|\mathbf{H}_{k} \cdot \mathbf{H}_l^H|^2$ can be expanded as the sum of $2^4=16$ terms as shown in Table~\ref{tab: breakdown}. Among them, many are conjugates of each other, including {\large \textcircled{\scriptsize 2}} and {\large \textcircled{\scriptsize 9}}, {\large \textcircled{\scriptsize 3}} and {\large \textcircled{\scriptsize 5}}, {\large \textcircled{\scriptsize 4}} and {\large \textcircled{\scriptsize 13}}, {\large \textcircled{\scriptsize 6}} and {\large \textcircled{\scriptsize 11}}, {\large \textcircled{\scriptsize 8}} and {\large \textcircled{\scriptsize 15}}, and {\large \textcircled{\scriptsize 12}} and {\large \textcircled{\scriptsize 14}}. It is easy to show that the sum of each pair has a zero mean. Thus, it remains to determine the rest four terms, as derived below.
 
 \begin{table}
\centering
\caption{Breakdown of \eqref{eq: HkHl-power}.}
\label{tab: breakdown}
\begin{tabular}{l||l }
\hline
\textcircled{\small{1}} $|\mathbf{H}_{k,\text{LOS}}\cdot \mathbf{H}^H_{l,\text{LOS}} |^2$ &  \textcircled{\small{2}} $\mathbf{H}_{k,\text{LOS}} \mathbf{H}^H_{l,\text{LOS}} \mathbf{H}_{l,\text{LOS}} \mathbf{H}^H_{k,\text{NLOS}}$  \\
\hline
\textcircled{\small{3}} $\mathbf{H}_{k,\text{LOS}} \mathbf{H}^H_{l,\text{LOS}} \mathbf{H}_{l,\text{NLOS}} \mathbf{H}^H_{k,\text{LOS}}$ & \textcircled{\footnotesize{4}} $\mathbf{H}_{k,\text{LOS}} \mathbf{H}^H_{l,\text{LOS}} \mathbf{H}_{l,\text{NLOS}} \mathbf{H}^H_{k,\text{NLOS}}$ \\
\hline
\textcircled{\small{5}} $\mathbf{H}_{k,\text{LOS}} \mathbf{H}^H_{l,\text{NLOS}} \mathbf{H}_{l ,\text{LOS}} \mathbf{H}^H_{k,\text{LOS}}$ & \textcircled{\small{6}} $\mathbf{H}_{k,\text{LOS}} \mathbf{H}^H_{l,\text{NLOS}} \mathbf{H}_{l,\text{LOS}} \mathbf{H}^H_{k,\text{NLOS}}$ \\
\hline
{\small \textcircled{\scriptsize 7}} $|\mathbf{H}_{k,\text{LOS}}\cdot \mathbf{H}^H_{l,\text{NLOS}} |^2$ & \textcircled{\small{8}} $\mathbf{H}_{k,\text{LOS}} \mathbf{H}^H_{l,\text{NLOS}} \mathbf{H}_{l,\text{NLOS}} \mathbf{H}^H_{k,\text{NLOS}}$ \\
\hline
\textcircled{\small{9}} $\mathbf{H}_{k,\text{NLOS}} \mathbf{H}^H_{l,\text{LOS}} \mathbf{H}_{l,\text{LOS}} \mathbf{H}^H_{k,\text{LOS}}$ & {\small \textcircled{\scriptsize 10}} $|\mathbf{H}_{k,\text{NLOS}}\cdot \mathbf{H}^H_{l,\text{LOS}} |^2$ \\
\hline
\textcircled{\small{11}} $\mathbf{H}_{k,\text{NLOS}} \mathbf{H}^H_{l,\text{LOS}} \mathbf{H}_{l,\text{NLOS}} \mathbf{H}^H_{k,\text{LOS}}$ & \textcircled{\small{12}} $\mathbf{H}_{k,\text{NLOS}} \mathbf{H}^H_{l,\text{LOS}} \mathbf{H}_{l,\text{NLOS}} \mathbf{H}^H_{k,\text{NLOS}}$ \\
\hline
\textcircled{\small{13}} $\mathbf{H}_{k,\text{NLOS}} \mathbf{H}^H_{l,\text{NLOS}} \mathbf{H}_{l,\text{LOS}} \mathbf{H}^H_{k,\text{LOS}}$ & \textcircled{\small{14}} $\mathbf{H}_{k,\text{NLOS}} \mathbf{H}^H_{l,\text{NLOS}} \mathbf{H}_{l,\text{LOS}} \mathbf{H}^H_{k,\text{NLOS}}$ \\
\hline
{\small \textcircled{\scriptsize 15}} $\mathbf{H}_{k,\text{NLOS}} \mathbf{H}^H_{l,\text{NLOS}} \mathbf{H}_{l,\text{NLOS}} \mathbf{H}^H_{k,\text{LOS}}$ & {\small \textcircled{\scriptsize 16}}  $|\mathbf{H}_{k,\text{NLOS}}\cdot \mathbf{H}^H_{l,\text{NLOS}} |^2$ \\
\hline
\end{tabular}
\end{table}

\begin{Thm}\label{Thm: 4}
The mean power of $\mathbf{H}_{k,\text{LOS}} \cdot \mathbf{H}_{l,\text{LOS}}^H$ ({\large \textcircled{\scriptsize 1}} in Table~\ref{tab: breakdown}) is given by
\begin{equation}\label{eq: mean-Hk_los-Hl_los-square}
\mathbb{E}[|\mathbf{H}_{k,\text{LOS}} \cdot \mathbf{H}_{l,\text{LOS}}^H|^2] = \frac{L_kL_l}{N^2} \left(\frac{\kappa}{\kappa+1}\right)^2. 
\end{equation}
\end{Thm}
\begin{IEEEproof} 
According to~\eqref{eq: Hk_los}, $\mathbf{H}_{k,\text{LOS}} \cdot \mathbf{H}_{l,\text{LOS}}^H$ can be expanded as
\begin{align}
\mathbf{H}_{k,\text{LOS}} \cdot \mathbf{H}_{l,\text{LOS}}^H &= \frac{1}{N^2}\frac{\kappa}{\kappa+1}\sum_{p=1}^{L_k} \alpha_{k,p} \bm{p}^T \mathbf{b}(\phi_1-\phi_{2,k,p}) \mathbf{a}^H(\phi^\text{AoD}) \mathbf{a}(\phi^\text{AoD}) \sum_{p'=1}^{L_l} \alpha^*_{l,p'} \mathbf{b}^H(\phi_1-\phi_{2,l,p'}) \bm{p}^*\nonumber \\
&= \frac{1}{N^2} \frac{\kappa}{\kappa+1} \sum_{p=1}^{L_k} \sum_{p'=1}^{L_l} \alpha_{k,p} \alpha^*_{l,p'} \bm{p}^T \mathbf{b}(\phi_1-\phi_{2,k,p}) \mathbf{b}^H(\phi_1-\phi_{2,l,p'}) \bm{p}^* 
\end{align}
using the fact that $\mathbf{a}^H(\phi^\text{AoD}) \mathbf{a}(\phi^\text{AoD}) = 1$. Hence, the mean power $\mathbb{E}[|\mathbf{H}_{k,\text{LOS}} \cdot \mathbf{H}_{l,\text{LOS}}^H|^2]$ can be expressed as
\begin{align}
\mathbb{E}[|\mathbf{H}_{k,\text{LOS}} \cdot \mathbf{H}_{l,\text{LOS}}^H|^2] &= \frac{1}{N^4} \Bigl(\frac{\kappa}{\kappa+1}\Bigr)^2 \mathbb{E}\Bigl[ \bm{p}^T\sum_{p=1}^{L_k} \sum_{p'=1}^{L_l} \alpha_{k,p} \alpha^*_{l,p'} \mathbf{b}(\phi_1-\phi_{2,k,p}) \mathbf{b}^H(\phi_1-\phi_{2,l,p'}) \bm{p}^* \nonumber \\
& \quad \times \bm{p}^T\sum_{q=1}^{L_k} \sum_{q'=1}^{L_l} \alpha_{l,q} \alpha^*_{k,q'} \mathbf{b}(\phi_1-\phi_{2,l,q}) \mathbf{b}^H(\phi_1-\phi_{2,k,q'}) \bm{p}^* \Bigr].
\end{align}
By grouping the terms associated with each user, we have
\begin{align}\label{eq: Hk_los-Hl_los_power}
\mathbb{E}[|\mathbf{H}_{k,\text{LOS}} \cdot \mathbf{H}_{l,\text{LOS}}^H|^2] &= \frac{1}{N^4}\Bigl(\frac{\kappa}{\kappa+1}\Bigr)^2  \mathbb{E}\Bigl[ \underbrace{ \sum_{p=1}^{L_k} \sum_{q'=1}^{L_l} \alpha_{k,p} \alpha^*_{k,q'} \bm{p}^T \mathbf{b}(\phi_1-\phi_{2,k,p}) \mathbf{b}^H(\phi_1-\phi_{2,k,q'}) \bm{p}^* }_{T_k} \Bigr]  \nonumber \\
& \quad \times \mathbb{E}\Bigl[ \underbrace{ \sum_{p'=1}^{L_l} \sum_{q=1}^{L_k} \alpha_{l,p'} \alpha^*_{l,q} \bm{p}^T \mathbf{b}(\phi_1-\phi_{2,l,p'}) \mathbf{b}^H(\phi_1-\phi_{2,l,q}) _{T_l} \bm{p}^*}_{T_l} \Bigr].
\end{align}
From \eqref{eq: alpha_square}, the independent fading of different paths implies that $T_k$ associated with the $k$-th UE can be reduced to
\begin{align}
\mathbb{E}[T_k] &=  \sum_{p=1}^{L_k} 
\mathbb{E}\Bigl[ |\bm{p}^T \mathbf{b}(\phi_1-\phi_{2,k,p})|^2 \Bigr] \nonumber \\
&= NL_k
\end{align}
where the second line is obtained from \eqref{eq: |p^Tb|^2}. 
Similarly, we can obtain $\mathbb{E}[T_l]= NL_l$. Substituting the obtained expressions for $\mathbb{E}[T_k]$ and $\mathbb{E}[T_l]$ into \eqref{eq: Hk_los-Hl_los_power}, we reach the desired result.
\end{IEEEproof}

\begin{Thm}\label{Thm: 5}
The mean power of $\mathbf{H}_{k,\text{LOS}} \cdot \mathbf{H}_{l,\text{NLOS}}^H$ ({\large \textcircled{\scriptsize 7}} in Table~\ref{tab: breakdown}) and that of $\mathbf{H}_{k,\text{NLOS}} \cdot \mathbf{H}_{l,\text{LOS}}^H$ ({\large \textcircled{\scriptsize 10}} in Table~\ref{tab: breakdown}) are identical, as given by
\begin{equation}\label{eq: mean-Hk_los-Hl_nlos-square}
\mathbb{E}[|\mathbf{H}_{k,\text{LOS}} \cdot \mathbf{H}_{l,\text{NLOS}}^H|^2] = \mathbb{E}[|\mathbf{H}_{k,\text{NLOS}} \cdot \mathbf{H}_{l,\text{LOS}}^H|^2] =\frac{L_k L_l}{N} \frac{\kappa}{(\kappa+1)^2}. 
\end{equation}
\end{Thm}
\begin{IEEEproof} 
We first rewrite $|\mathbf{H}_{k,\text{LOS}} \cdot \mathbf{H}_{l,\text{NLOS}}^H|^2$ as $\mathbf{H}_{k,\text{LOS}} \mathbf{H}^H_{l,\text{NLOS}} \mathbf{H}_{l,\text{NLOS}} \mathbf{H}^H_{k,\text{LOS}}$ and thus,
\begin{equation}\label{eq: mean-Hk_los-Hl_nlos}
\mathbb{E}[|\mathbf{H}_{k,\text{LOS}} \cdot \mathbf{H}_{l,\text{NLOS}}^H|^2] = \mathbb{E}_{\mathbf{H}_{k,\text{LOS}}} \Bigl[\mathbf{H}_{k,\text{LOS}} \cdot \mathbb{E}_{\mathbf{H}_{l,\text{NLOS}}}[\mathbf{H}^H_{l,\text{NLOS}} \cdot \mathbf{H}_{l,\text{NLOS}}] \cdot \mathbf{H}^H_{k,\text{LOS}}\Bigr].
\end{equation}
According to \eqref{eq: H_nlos}, the inner expectation can be expressed as
\begin{align}\label{eq: mean-H_nlos^H-H_nlos}
\mathbb{E}_{\mathbf{H}_{l,\text{NLOS}}}[\mathbf{H}^H_{l,\text{NLOS}} \cdot \mathbf{H}_{l,\text{NLOS}}] = \frac{1}{\kappa+1} \mathbb{E}_{\mathbf{G}_\text{NLOS}, \mathbf{h}_l}[ \|\mathbf{G}^H_\text{NLOS} \text{diag}( \mathbf{h}^*_l )\bm{p}^* \|^2 ].
\end{align}
From~\eqref{eq: ris-user channel-v2}, $\text{diag}( \mathbf{h}^*_l )\bm{p}^*$ is an $N \times 1$ vector given by
\begin{equation}
\text{diag}( \mathbf{h}_l^* )\bm{p}^* =  \frac{1}{\sqrt{N}}\Bigl[ \sum_{p=1}^{L_l} \alpha^*_{l,p} e^{j(a_{l,p,1}-\varphi_1)},\cdots,\sum_{p=1}^{L_l} \alpha^*_{l,p} e^{j(a_{l,p,N}-\varphi_N)} \Bigr]^T
\end{equation}
where $a_{l,p,n_1\cdot N_2 +n_2+1}=2\pi(n_1\cdot \phi_{2,l,p} + n_2 \cdot \vartheta_{2,l,p})$ for $n_1=\{0,1,\cdots,N_1-1\}$ and $n_2=\{0,1,\cdots,N_2-1\}$. Then $\mathbf{H}^H_{l,\text{NLOS}} \in \mathbb{C}^M$ in \eqref{eq: mean-H_nlos^H-H_nlos} can be written as
\begin{equation}\label{eq: H_lnlos_expan}
\mathbf{H}^H_{l,\text{NLOS}}=\frac{1}{\sqrt{\kappa+1}}\mathbf{G}^H_\text{NLOS} \text{diag}( \mathbf{h}^*_l )\bm{p}^* = \frac{1}{\sqrt{N(\kappa+1)}}\Bigl[ \sum_{n=1}^N g^*_{n,1} \sum_{p=1}^{L_l} \alpha^*_{l,p} e^{j(a_{l,p,n}-\varphi_n)}, \cdots, \sum_{n=1}^N g^*_{n,M} \sum_{p=1}^{L_l} \alpha^*_{l,p} e^{j(a_{l,p,n}-\varphi_n)} \Bigr]^T.
\end{equation}
Accordingly, $\mathbf{H}^H_{l,\text{NLOS}}\mathbf{H}_{l,\text{NLOS}}$ is an $M\times M$ matrix and its $(i,j)$-th entry, for $i=1,\cdots,M$ and $j=1,\cdots,M$, is given by
\begin{align}\label{eq: E[eq: H_lnlos_expan]}
\mathbf{H}^H_{l,\text{NLOS}} \mathbf{H}_{l,\text{NLOS}}(i,j) = \frac{1}{N(\kappa+1)} \sum_{n=1}^N \sum_{n'=1}^N g_{n,i}^* g_{n',j} \sum_{p=1}^{L_l} \sum_{p'=1}^{L_l} \alpha_{k,p}^* \alpha_{l,p'}e^{j(a_{l,p,n}-a_{l,p,n'})+j(\varphi_{n'}-\varphi_n)}.
\end{align}
According to \eqref{eq: alpha_square} and \eqref{eq: g_square}, the independent fading assumption of different multipath components suggests that the $(i,j)$-th entry of $\mathbb{E}[\mathbf{H}^H_{l,\text{NLOS}}\mathbf{H}_{l,\text{NLOS}}]$ is zero when $i \neq j$, i.e., the off-diagonal entry. For $i=j$, \eqref{eq: E[eq: H_lnlos_expan]} reduces to
\begin{equation}
\mathbb{E}[\mathbf{H}^H_{l,\text{NLOS}}\mathbf{H}_{l,\text{NLOS}}(i,i)] = \begin{cases}
\frac{L_l}{N(\kappa+1)}, & n=n'~\text{and}~p=p', \\
0, & \text{otherwise}.
\end{cases}
\end{equation}
As a result, $\mathbb{E}[\mathbf{H}^H_{l,\text{NLOS}}\mathbf{H}_{l,\text{NLOS}}]=\frac{L_l}{N(\kappa+1)}\mathbf{I}_M$. Together with~\eqref{eq: Hk_los}, \eqref{eq: mean-Hk_los-Hl_nlos} can be arranged as
\begin{align}
\mathbb{E}[|\mathbf{H}_{k,\text{LOS}}\mathbf{H}_{l,\text{NLOS}}^H|^2] &= \frac{\kappa L_l}{N^2(\kappa+1)^2} \mathbb{E} \Bigl[ \sum_{p=1}^{N_k} \sum_{p'=1}^{N_k} \alpha_{k,p} \alpha^*_{k,p'} \bm{p}^T \mathbf{b}^H(\phi_1-\phi_{2,k,p}, \varphi_1-\varphi_{2,k,p'}) \mathbf{a}^H(\phi^\text{AoD}) \mathbf{a}(\phi^\text{AoD})  \nonumber \\
  &\times \mathbf{I}_M\mathbf{b}(\phi_1-\phi_{2,k,p'}, \varphi_1-\varphi_{2,k,p'}) \bm{p}^* ] \nonumber \\
  &= \frac{\kappa L_k}{N^2(\kappa+1)^2} \sum_{p=1}^{L_k} \mathbb{E}[ |\bm{p}^T \mathbf{b}(\phi_1-\phi_{2,k,p}, \varphi_1-\varphi_{2,k,p'}) |^2 ]. 
\end{align}
Again, $\mathbf{a}^H(\phi^\text{AoD}) \mathbf{I}_M \mathbf{a}(\phi^\text{AoD})=1$ and using \eqref{eq: |p^Tb|^2}, we obtain~\eqref{eq: mean-Hk_los-Hl_nlos-square}. As to $\mathbf{H}_{k,\text{NLOS}} \cdot \mathbf{H}_{l,\text{LOS}}^H$, it has a symmetric structure as $\mathbf{H}_{k,\text{LOS}} \cdot \mathbf{H}_{l,\text{NLOS}}^H$ and thus the same mean power. This completes the proof.
\end{IEEEproof}

\begin{Thm}\label{Thm: 6}
The mean power of $\mathbf{H}_{k,\text{NLOS}} \cdot \mathbf{H}_{l,\text{NLOS}}^H$ ({\large \textcircled{\scriptsize 16}} in Table~\ref{tab: breakdown}) given by
\begin{equation}\label{eq: mean-Hk_los-Hl_nlos'-square}
\mathbb{E}[|\mathbf{H}_{k,\text{NLOS}}\mathbf{H}_{l,\text{NLOS}}^H|^2] = M\frac{L_kL_l}{(\kappa+1)^2}. 
\end{equation}
\end{Thm}
\begin{IEEEproof} 
Using\eqref{eq: H_lnlos_expan}, the inner product between $\mathbf{H}_{k,\text{LOS}}$ and $\mathbf{H}_{l,\text{LOS}}$ can be expressed as
\begin{equation}
\mathbf{H}_{k,\text{LOS}} \cdot \mathbf{H}^H_{l,\text{LOS}} = \frac{1}{N(\kappa+1)}(x_1+\cdots+x_M)
\end{equation}
where 
\begin{align}
x_m =& \sum_{n=1}^N \sum_{p=1}^{L_k} \sum_{n'=1}^N \sum_{p'=1}^{L_l} \sum_{m=1}^N \sum_{q=1}^{L_k} \sum_{m'=1}^N \sum_{q'=1}^{L_l}  \alpha_{k,p} \alpha^*_{k,q} \alpha_{l,p'} \alpha^*_{l,q'} h_{n,1} h^*_{m,1} g^*_{n',1} g_{m',1} \nonumber \\
& \times e^{j(\vartheta_n-a_{k,p,n}+a_{l,p',n'} - \vartheta_{n'})} e^{-j(\vartheta_m-a_{k,q,m}+a_{l,q',m'} - \vartheta_{m'})},~m=1,\cdots,M.
\end{align}
Then the mean power of $\mathbf{H}_{k,\text{NLOS}} \cdot \mathbf{H}_{l,\text{NLOS}}^H$ is given by
\begin{align}\label{eq: mean-Hk_los-Hl_nlos'-square_v2}
\mathbb{E}[|\mathbf{H}_{k,\text{NLOS}} \cdot \mathbf{H}_{l,\text{NLOS}}^H|^2] &= \frac{1}{N^2(\kappa+1)^2} \mathbb{E}[(x_1+\cdots+x_M)\cdot (x^*_1+\cdots+x^*_M)].
\end{align}
Following \eqref{eq: alpha_square} and \eqref{eq: g_square}, it is readily obtained that
\begin{equation}\label{eq: mean-xi-times-xi'}
\mathbb{E}[x_i\cdot x^*_{i'}] = \begin{cases} 
\frac{L_k L_l}{(\kappa+1)^2}, &n=m, p=q, n'=m', p'=q', \\
0, &~\text{otherwise},
\end{cases}
\end{equation}
for $i=1,\cdots, M$. Substituting \eqref{eq: mean-xi-times-xi'} into \eqref{eq: mean-Hk_los-Hl_nlos'-square_v2}, we obtain \eqref{eq: mean-Hk_los-Hl_nlos'-square} that completes the proof.
\end{IEEEproof}

\subsection{Approximated and Asymptotic Mean Correlation Coefficient}

Based on the above results and the approximated expression for the mean correlation coefficient in \eqref{eq: H-square}, we obtain the approximated mean correlation coefficient between two cascade channels and analyze its convergence in the following.

\begin{Thm}\label{thm: 7}
The approximated mean squared correlation coefficient between two cascade RIS channels is given by
\begin{equation}\label{eq: mean-rho^2}
\mathbb{E}[\rho_{i,j}^2] \approx 1-\frac{ (M-1)(M+2\frac{\kappa}{N}) }{ (\frac{\kappa}{N}+M)^2 }.
\end{equation}
\end{Thm}
\begin{IEEEproof} 
The result is obtained by substituting \eqref{eq: H_los_square_final}, \eqref{eq: H_nlos_square_final}, \eqref{eq: mean-Hk_los-Hl_los-square}, \eqref{eq: mean-Hk_los-Hl_nlos-square} and \eqref{eq: mean-Hk_los-Hl_nlos'-square} into \eqref{eq: approx-mean-squared-corr-coeff-indep}. 
\end{IEEEproof}
Theorem~\ref{thm: 7} indicates that the mean correlation coefficient is a function of the Rician factor, the number of BS antennas and the number of RIS elements; it is independent from the number of paths in the BS-RIS channel and the phase configuration on the RIS.
We further consider the extreme case when the RIS has a massive number of elements.
\begin{Cor}\label{cor: 1}
When $N\rightarrow \infty$, the mean correlation coefficient between two cascade RIS channels converges to a constant as given by
\begin{equation}\label{eq: corr-coeff-M-large-final}
\lim_{N \rightarrow \infty} \mathbb{E}[ \rho_{i,j} ] \rightarrow \sqrt{\frac{1}{M}}.
\end{equation}
The same holds when $\kappa \rightarrow 0$, i.e., the BS-RIS channel experiences Rayleigh fading. 
\end{Cor}
\begin{IEEEproof}
When $N \rightarrow \infty$ or $\kappa \rightarrow 0$, $\kappa/N \rightarrow 0$ and thus, $ \mathbb{E}[ \rho_{i,j}^2 ]$ in \eqref{eq: mean-rho^2} is asymptotically equal to $1/M$. By taking the square root of $ \mathbb{E}[ \rho_{i,j}^2 ]$, we obtain the desired result.
\end{IEEEproof}
Corollary~\ref{cor: 1} suggests that when there is no LoS in the BS-RIS channel, the correlation between two cascade RIS channels only depends on the number of BS antennas but not on the number of RIS elements. As a result, the correlation between two cascade RIS channels can be reduced by a large RIS with many elements only when there is a LoS in the BS-RIS channel. This emphasizes the importance of properly deploying the RIS to have the LoS path in the BS-RIS channel. 

%
\begin{Cor}\label{cor: 2}
\begin{itemize}
\item When $M=1$, $\mathbb{E}[\rho_{i,j}]=1$. 
\item When the BS-RIS channel is LoS-dominant (i.e., $\kappa$ is sufficiently large) such that $\kappa/N \gg M$, $\mathbb{E}[ \rho_{i,j} ]$ is an increasing function of $\kappa$ and 
\begin{equation}\label{eq: rho_k/N >> M}
\lim_{\kappa/N \gg M} \mathbb{E}[ \rho_{i,j} ] \rightarrow 1.
\end{equation}
\end{itemize}
\end{Cor}
\begin{IEEEproof}
\eqref{eq: rho_k/N >> M} can be readily verified from \eqref{eq: mean-rho^2}.
\end{IEEEproof}
Corollary~\ref{cor: 2} suggests when the BS has only one antenna, the correlation between two cascade channels is irrelevant to the number of RIS elements. Also, if the BS-RIS channel shared by different users is dominated by the LoS path, two cascade channels through a common RIS may have non-negligible correlation. For example, when $M=2$, $N=25$, and $\kappa=12$ dB, the mean correlation coefficient is about 0.53.

\section{Numerical Results}\label{sec: results}

We validate the analysis accuracy through simulations. The RIS has a UPA with an equal number of elements in each dimension and 10,000 channel realizations are considered. The simulated mean correlation coefficient $\mathbb{E}[\rho]$ and the mean squared correlation coefficient $\mathbb{E}[\rho^2]$ are obtained based on \eqref{eq: corr-coeff-def} and \eqref{eq: approx-mean-squared-corr-coeff-dependence}, respectively. Their theoretical values are obtained using~\eqref{eq: approx-mean-squared-corr-coeff-indep} and then \eqref{eq: ub-mean-corr-coeff}, respectively. Unless specified, we set the number of BS antennas $M=2$, the Rician factor $\kappa=5$, and $d_B=d_R=\lambda/2$. 

\begin{figure}[!t]
    \subfloat[Equal phase.\label{fig: 2a}]{%
      \includegraphics[width=0.5\textwidth]{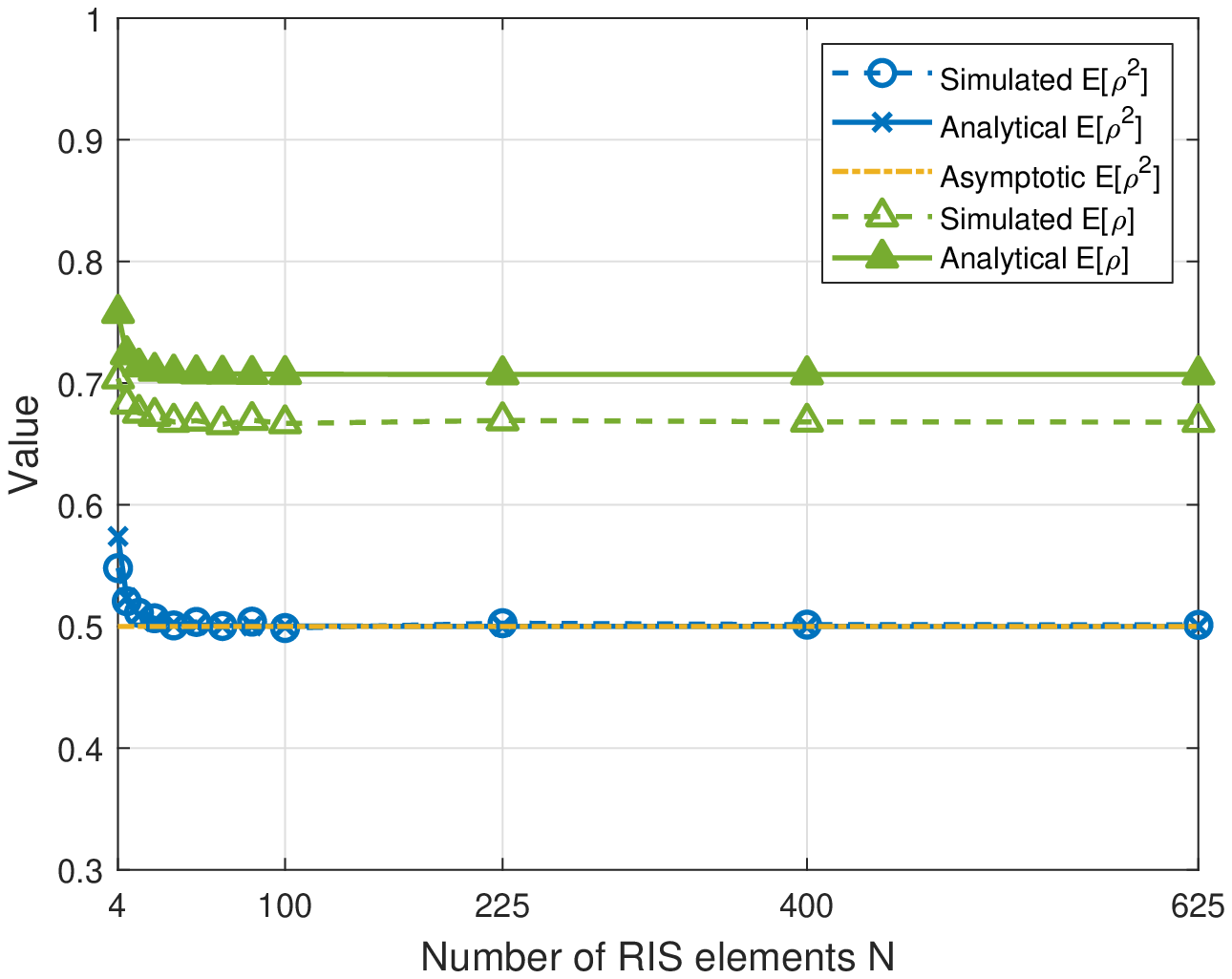}
    }
    \hfill
    \subfloat[Unequal phase.\label{fig: 2b}]{%
      \includegraphics[width=0.5\textwidth]{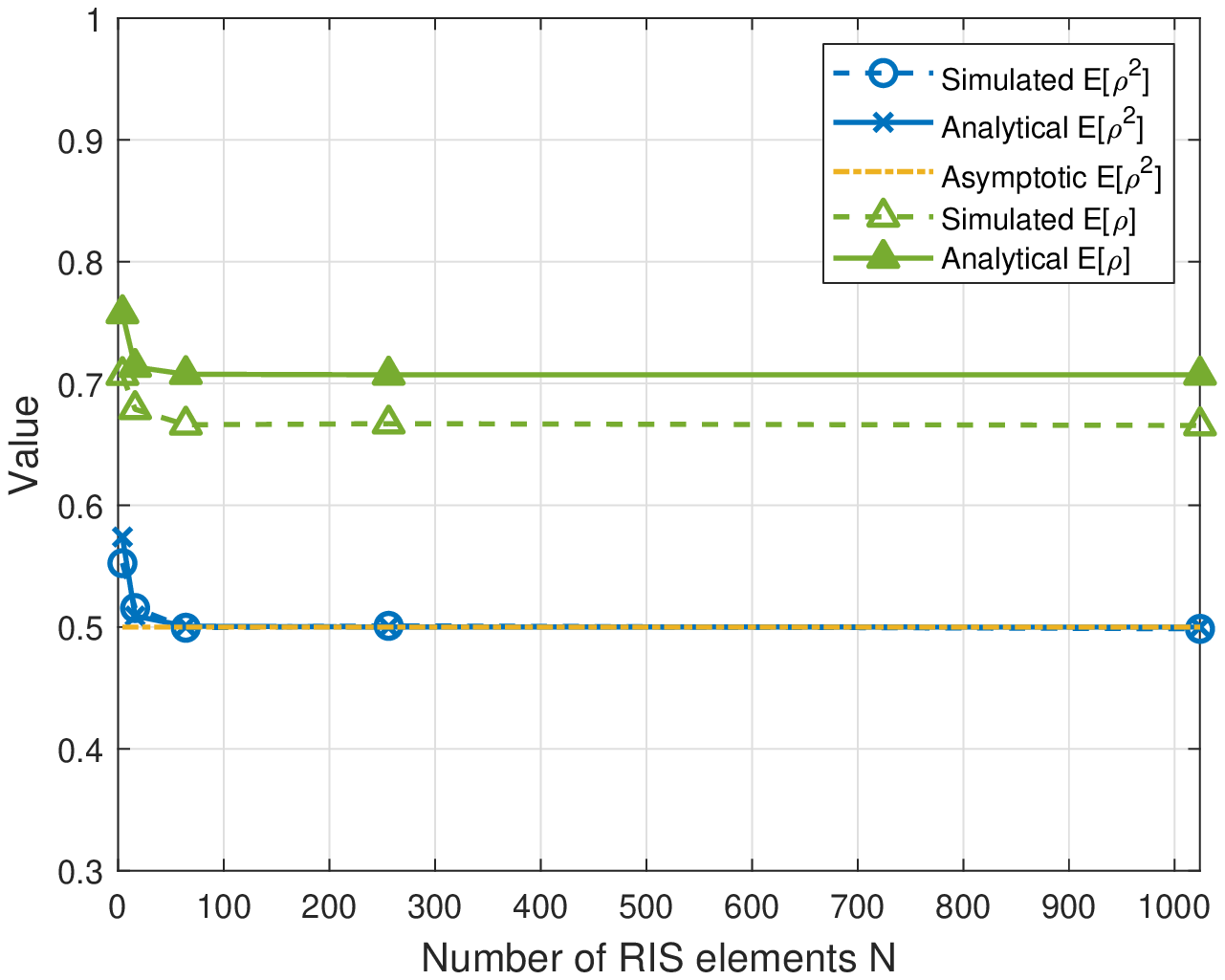}
    }
    \caption{The mean squared correlation coefficient and the mean correlation coefficient vs. $N$ for $M=2$ BS antennas.}
    \label{fig: corr-vs-N}
  \end{figure}

Since the mean correlation coefficient between two cascade channels is obtained by first computing the squared value, we show examine the analytical and simulated mean squared correlation coefficients. Here, two different phase configurations for the RIS are considered. In Fig.~\ref{fig: 2a}, all RIS elements are assigned with an equal phase $\varphi_n=\pi/6$. In Fig.~\ref{fig: 2b}, unequal phases are applied to RIS elements based on the binary-tree based codebook proposed in~\cite{Xiao2016}, which has been widely considered for mmWave communications in the literature. The codebook contains $k=\log_2(N)$ layers where $N$ is the number of antennas under consideration and each layer has $2^k$ codewords to generate varied steering angles and beam coverage. For our purpose, $N$ equals the number of RIS elements on the UPA. Among all the codewords, the one provides the maximum summed channel power of the two cascade channels is used to determine the phase of each RIS element. It can be seen from both Fig.~\ref{fig: 2a} and Fig.~\ref{fig: 2b} that the analytical mean squared correlation coefficients closely match to the simulated ones, indicating the validity of the approximation made in~\eqref{eq: approx-mean-squared-corr-coeff-indep}. As shown in the figure, the mean squared correlation coefficient decreases with $N$ and it converges to $1/M=0.5$ when $N$ is sufficiently large, confirming Corollary~\ref{cor: 1}. 
An RIS with a large number of is likely to be of practical interest, in which case our analysis provides convincing results. In our setting, the mean squared correlation coefficient is nearly converged when $N=400$. By taking the square root of the mean squared correlation coefficient, the analytical mean correlation coefficient remains close to the simulated one  yet their gap does not diminish as $N$ increases, due to the Jensen's inequality in \eqref{eq: ub-mean-corr-coeff}. Comparing~Fig.~\ref{fig: 2a} and Fig.~\ref{fig: 2b}, the equal phase configuration has the same mean correlation coefficients as the non-equal phase configuration. It is thus concluded that the mean correlation coefficient between two cascade channels depends on the amount but not the phase values of RIS elements.

\begin{figure}[!t]
\centering
{
\includegraphics[width=0.75\linewidth]{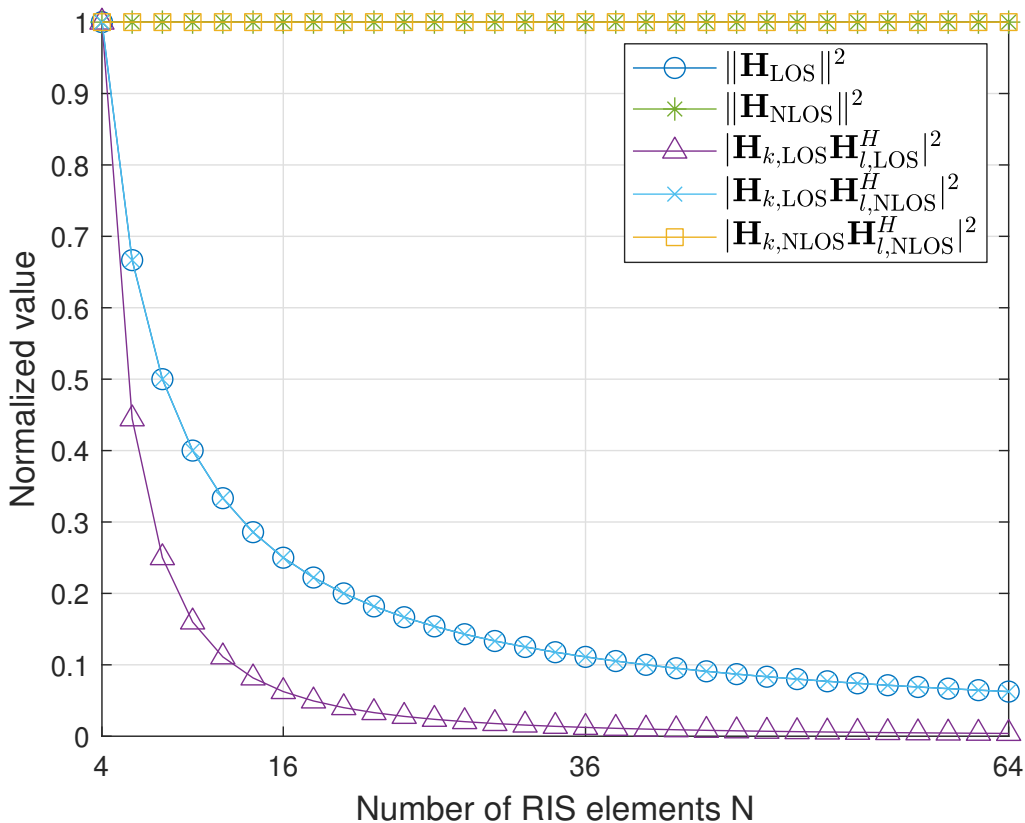}
}
\caption{Comparison of the normalized mean power of various channel components as a function of $N$.}
\label{fig: mean-power-component}
\end{figure}

To understand how the number of RIS elements $N$ changes the power of individual channel component and in turn affects the channel correlation, Fig.~\ref{fig: mean-power-component} plots the mean power of different channel components relevant to the correlation coefficient. To ease the illustration, only the theoretical values are shown and all the mean power values are normalized to one. On can see that $\mathbb{E}[\|\mathbf{H}_{\text{NLOS}} \|^2]$ and $\mathbb{E}[|\mathbf{H}_{k,\text{NLOS}} \mathbf{H}^H_{l,\text{NLOS}}|^2]$ that are purely determined by the NLoS paths are invariant to $N$. This reveals that the NLoS power in the user-independent channel between BS and RIS is not affected by $N$ and thus does not change the correlation between two cascade channels. This also implies that for two users, if the commonly shared BS-RIS channel contains NLoS paths only (i.e., $\kappa=0$), their mean correlation coefficient is constant to $N$, which justifies Corollary~\ref{cor: 1}. On the other hand, the curves of $\mathbb{E}[\|\mathbf{H}_\text{LOS}\|^2]$ and $\mathbb{E}[|\mathbf{H}_{k,\text{LOS}} \mathbf{H}^H_{l,\text{NLOS}}\|^2]$ have the same decaying rate equal to $1/N$ that can be also seen from \eqref{eq: H_los_square_final} and \eqref{eq: mean-Hk_los-Hl_nlos-square}. Finally, $\mathbb{E}[|\mathbf{H}_{k,\text{LOS}} \mathbf{H}^H_{l,\text{LOS}}|^2]$ also decays with $N$ but in a higher rate equal to $1/N^2$, according to \eqref{eq: mean-Hk_los-Hl_los-square}. The above result indicates that the LoS strength in the BS-RIS channel plays the major role in determining the correlation between two cascade channels.
 
\begin{figure}[!t]
\centering
{
\includegraphics[width=0.75\linewidth]{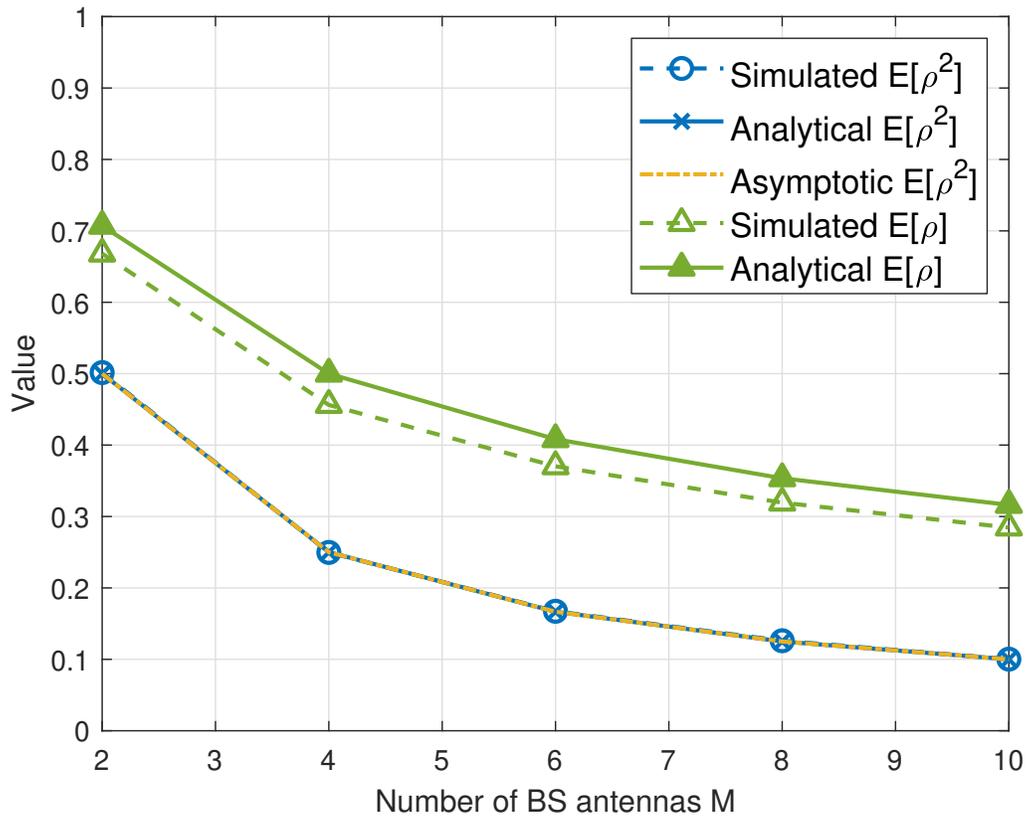}
}
\caption{The mean squared correlation coefficient and the mean correlation coefficient vs. $M$ for $N=400$ RIS elements.}
\label{fig: corr-vs-M}
\end{figure}

Fig.~\ref{fig: corr-vs-M} shows the impact of the number of BS antennas to the channel correlation for $N=400$ RIS elements each with an equal phase of $6\pi$. Again, the analytical mean squared correlation coefficients agree with the simulated ones, indicating the tightness of the approximation in \eqref{eq: approx-mean-squared-corr-coeff-indep}. Also, the mean squared correlation coefficient converges to the asymptotic value given in Corollary~\ref{cor: 1}. Similar to Fig.~\ref{fig: corr-vs-N}, there exists a gap between the simulated mean correlation coefficient and the theoretical ones, due to the approximation in \eqref{eq: ub-mean-corr-coeff}. 

Finally, we investigate the impact of LoS dominance to the correlation between two cascade channels by varying the Rician factor from $\kappa=0$ to $\kappa=10$ for $N=4$ and 64. The range of $\kappa$ is selected according to the measurements for 28 GHz mmWave channels~\cite{Samimi2016}, where the typical value for $\kappa$ falls in the range of $\kappa \in [3,10]$ dB. From the figure, the correlation between two cascade RIS channels is affected differently by $\kappa$ depending on the number of RIS elements. Specifically, the mean squared correlation coefficient increases with $\kappa$ for a small $N$ (e.g., $N=4$) but it becomes invariant to $\kappa$ for a greater $N$ (e.g., $N=64$). This can be explained by Corollary~\ref{cor: 2}, which reveals that the LoS path in the BS-RIS channel affects the average correlation coefficient only when the number of RIS elements is small such that $\kappa/N \gg M$. It is noticed that when $N=4$, the analytical value is deviated from the simulated one as $\kappa$ increases. The discrepancy is mainly caused by the ignored dependence between the numerator and the denominator in \eqref{eq: approx-mean-squared-corr-coeff-indep}. The approximation is tight when $N$ is large, which would likely be the case in practice.

\begin{figure}[!t]
\centering
{
\includegraphics[width=0.75\linewidth]{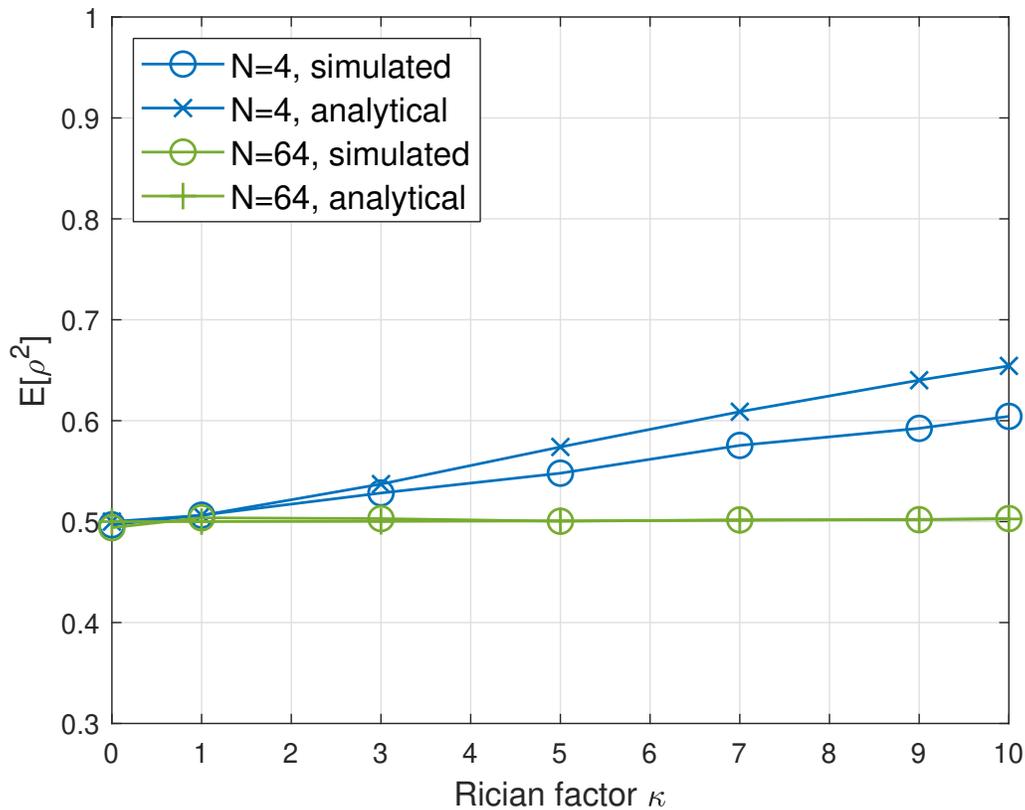}
}
\caption{The mean squared correlation coefficient vs. Rician factor $\kappa$.}
\label{fig: corr-vs-rician}
\end{figure}

\section{Conclusion}\label{sec: conclusion}
This work theoretically analyzes the correlation between two cascade channels in the RIS-aided communication system using the mean correlation coefficient between two channel vectors as the metric. Simulation results show that our analysis is accurate for a wide range of system parameters. It its observed that the mean correlation coefficient of the cascade channels heavily depends on the channel characteristics of the BS-RIS channel shared by different users. In the presence of the LoS path in the BS-RIS channel, the mean correlation coefficient decays as the number of RIS elements $N$ increases and it converges to $\sqrt{1/M}$ quickly where $M$ is the number of BS antennas. Also, the mean correlation coefficient does not change with the phase configuration on the RIS. In practice, the LoS path in the BS-RIS channel may be favored considering the passive nature of the RIS. In this case, it is beneficial to employ a sufficiently large antenna array at the BS to avoid strong correlation between two users sharing the same RIS, while the number of RIS elements does not need to be very large (greater than 100).

\bibliographystyle{IEEEtran}
\bibliography{IEEEabrv,ref_new}


 




\vfill

\end{document}